# Experimental Observation of Time-Domain Bound States in The Continuum

**Zahra Manzoor[1]†, Oded Schiller[2,4]†, Yonatan Plotnik[3,4], Mordechai Segev[2,3,4]\* and Dimitrios Peroulis[1]**

[1]*ECE Department, Purdue University; West Lafayette, IN 47907, USA*
[2]*Electrical and Computer Engineering Department, Technion; Haifa, Israel*
[3]*Physics Department, Technion - Israel Institute of Technology; Haifa, Israel*
[4]*Solid-State Institute, Technion - Israel Institute of Technology; Haifa, Israel.*
† *These authors contributed equally to this work*

*\*Corresponding author. Email: msegev@technion.ac.il*

## Abstract:

Bound states in the continuum (BICs) are spatially localized eigenmodes that remain perfectly confined even though their energies reside within a continuum of radiating modes. BICs were predicted in 1929, but their experimental realization awaited more than 8 decades. Following their experimental observation, BICs were explored in a variety of wave systems, and found to exhibit a plethora of fundamental features such nontrivial topology and extremely high Q-factor. Recently, with foundational advances in the new field of electromagnetic waves in time-varying media, BICs were predicted to exist in the time domain, with their wavenumber embedded in a continuum of unbound momentum modes. Here, we present the first experimental realization of the time-domain Bound States in the Continuum. We use a transmission-line network with a time-modulated wave-impedance and show that a sinusoidal wave launched into the network naturally evolves into a time-domain BIC with a well-defined peak and decaying-oscillating tails. We show that the time-domain BIC is anti-symmetric despite the symmetric nature of the modulation. These experiments pave the way for exploring new phenomena in the fields of BICs and time-varying wave-systems in nonconservative regimes where time-translation symmetry is broken.

## Main Text:

Bound state in the continuum (BIC) is a localized eigenstate embedded in the continuum of unbounded modes. BIC modes where first conceived in the context of quantum mechanics in 1929 by John Von Neumann and Eugene Wigner (*1*). Generally, a quantum system with a finite potential well, which is described by the Schrödinger equation, exhibits two kinds of eigenstates. The first kind is a finite number of bound states: square integrable states that are localized inside the potential well, and whose energies lie below the ambient level of the potential. The second type of modes are the continuum of unbounded modes, with energies that lie above the ambient level of potential. In their groundbreaking paper, Von Neumann and Wigner constructed a special kind of a finite potential well, which has one eigen state with energy above the continuum threshold, but counterintuitively the mode is square integrable and localized at the position of the potential well. Over the years, the concept of BICs was extended to many kinds of wave systems, and theoretically BIC states were identified in electromagnetic (EM) waves (*2–6*), sound waves (*7, 8*) and other quantum systems (*9–13*).

One can also think of BIC as a sort of resonant state, where the lifetime of the resonant state becomes infinite. A resonant state is a partially-localized state with a finite lifetime, that decays as it couples to continuum modes (*9, 14–18*). The decay rate dictates for how long the mode stays localized. In some special cases, the coupling of the resonant modes to the radiation modes can be nulled, usually because of destructive interference between different loss channels. In these cases, the decay rate goes to zero and the lifetime becomes infinite; these are exactly BIC states. The fact that BIC modes have extremely high (ideally infinite) Q-factor, means that they can greatly enhance light-matter interactions, for example - give rise to ultralow threshold lasers (*19*).

For decades, BICs were only a theoretical concept, stemming only from mathematical curiosity. The main reason BICs could not be observed for many decades can be understood by looking at Von Neumann and Wigner original BIC. The potential structure that supports the BIC, although decaying, has infinite support in space, namely, the potential is decaying in an oscillating fashion *ad infinitum*. When trimming this potential, which will always happen in any physically realizable setting, the BIC state immediately couples to the continuum modes, and does not stay localized. This issue has been hampering theoretical suggestions for many years. Early pioneering attempts to observe experimentally BIC-like states, such as the observation of bound modes above the potential well([20]), ended up as being defect modes: when taking the continuum limit and creating an infinite system, that mode disappears. The first experimental observation of BIC state happened in 2011 ([21]), for paraxial electromagnetic (EM) waves. This experiment took advantage of the realization that some potentials with special kinds of symmetry can support a BIC that will not couple to continuum modes even when the system is finite. The experiment created an array of paraxial waveguides, with a specific symmetry that supports a BIC. After that first experiment, BIC states were observed in many more wave systems([22–29]). Furthermore, later theoretical works showed that BICs can have interesting topological properties([30–33]).

Recently, with the surge of interest in wave in time-varying media ([34–45]) and specifically Photonic Time-Crystals ([46–51]) and the realization that wave systems can exhibit momentum gaps, a new idea emerged: time-domain bound states in the continuum ([52]). Time-domain BIC are modes that are localized (square integrable) in time and embedded in the continuum of wavenumbers of unbounded temporal modes. The time-domain BIC was theorized by envisioning a time-varying dielectric function

$\varepsilon(t)$, that supports a localized mode with a specific wavenumber, $k_{BIC}$. At $t \to -\infty$ the mode has zero energy, but it rises with time until forming a peak in the vicinity of $t = 0$, by drawing energy from the temporal modulation of $\varepsilon(t)$, and subsequently decays to zero for $t \to \infty$, as the modulation is drawing energy back from the state. The dielectric function is chosen such that all other modes with wavenumbers $k \neq k_{BIC}$ are unbounded (not square integrable) when $|t| \to \infty$, making this state a true time-domain BIC.

**Here, we present the first experimental realization of time-domain Bound states in the Continuum.** We employ a compact, time-modulated transmission-line network in which the time-varying permittivity ε(t) of a medium, first introduced theoretically in (*52*), is mapped onto a modulated capacitance, C(t), within a distributed LC ladder circuit. We show that, for a specific input frequency, a simple sinusoidal wave launched into the network evolves into a temporally localized mode. By slightly changing the input frequency, we show that the mode stops being localized, as expected from a BIC. We also confirm the theoretical prediction in (*52*) of the symmetry of the time-domain BIC, showing that time-domain BIC eigenmode is necessary anti-symmetric in time, even though the modulation is symmetric in time. This first experimental observation of time-domain BICs fundamentally extends the physics of wave propagation in time-varying media and paves the way new applications of BICs.

We realize BICs in a radio frequency (RF) networks, where we map the time-varying permittivity ε(t) (of (*52*)) onto modulated capacitance C(t) in the network shown in Fig. 1. Inductors reproduce the magnetic permeability of the medium, $\mu$, while the varactor diodes control its time-dependent capacitance. Controlled modulation of the capacitance creates an effective time-varying medium whose impedance varies in time according to $Z(t) \propto \sqrt{\frac{1}{C(t)}} = \sqrt{\frac{1}{\varepsilon(t)}}$. When the modulation parameters satisfy the BIC condition, the temporal modulation confines the output waveform within a finite temporal window - forming time-domain BIC. As we show below, launching a sinusoidal wave at the predesigned frequency into our networks results in a localized peak in time, while small deviations from that frequency destroy the BIC by becoming a delocalized temporal mode. Detailed analysis of our experimental platform, along with its modeling and simulations, are provided in the Supplementary Information. This system provides an experimentally accessible platform that reproduces the theoretical wave equation for time-domain BICs, offering tunable control over confined, radiative, and divergent regimes.

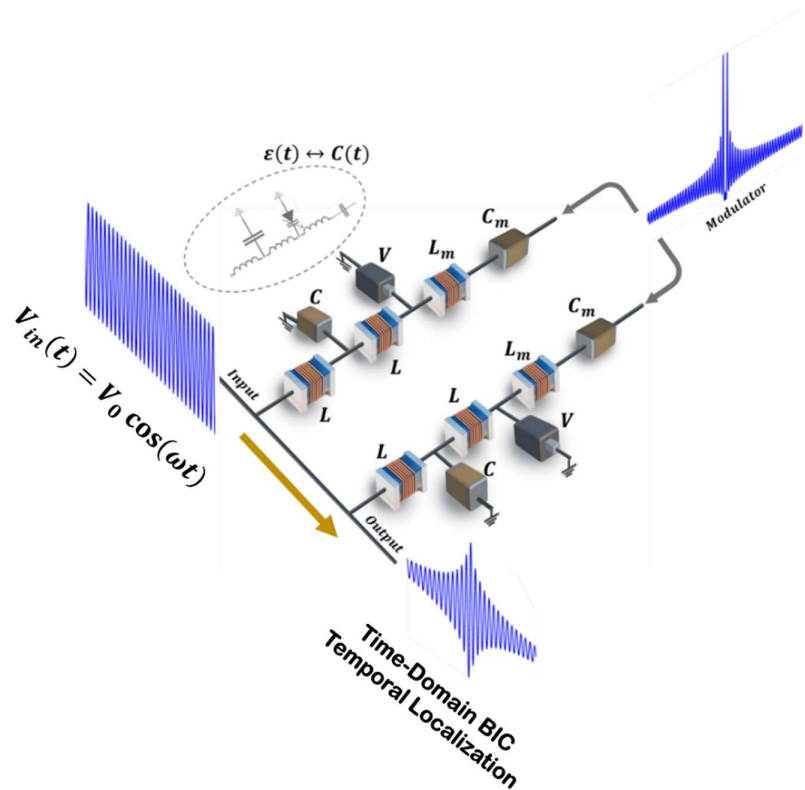

**Fig. 1. Illustration of the Experimental set up for generation of time-domain BICs.** The periodic input signal enters a time-varying transmission line, where the C, L components represent the capacitance and inductance of the transmission line. The capacitance of the transmission line is time-varying because of the inclusion of varactors, the Var component in the illustration. The time-varying capacitance is modulated such that the transmission line transforms a sinusoidal input signal into a localized time-domain BIC, if the signal is at the correct (predesigned) frequency.

The experimental configuration required precise synchronization between the 200-300 MHz modulation, which is used to modulate the capacitance of the varactor, and the 60–100 MHz input signal used to probe the temporal response of the system and generate the BIC. Both signals were generated from the same reference clock to eliminate phase drift and ensure deterministic timing between the launched signal and the modulation envelope. The BIC is generated by launching a sinusoidal signal into the circuit. Figure 2 shows typical experimental results. The BIC forms at a specific modulation frequency, in our case – 62 MHz, displaying a clear anti-symmetric peak in time, with a small pedestal of amplitude ~50 times smaller, Fig. 2b. The pedestal arises from the fact that the modulation is finite in time, and from the resistance of the varactor, which perturbs the effective wave impedance, as well as from residual RF signal leakage from the main signal path into the varactor. The time-domain BIC persists over multiple oscillation cycles before slowly decaying. Deviations from the BIC frequency lead to a smaller peak and an larger pedestal, accompanied by faster decay - signatures of partial coupling to the continuum, Fig. 2a and 2c. These trends confirm that the BIC appears at a discrete frequency of 62 MHz. For further deviations from this frequency the peak completely vanishes.

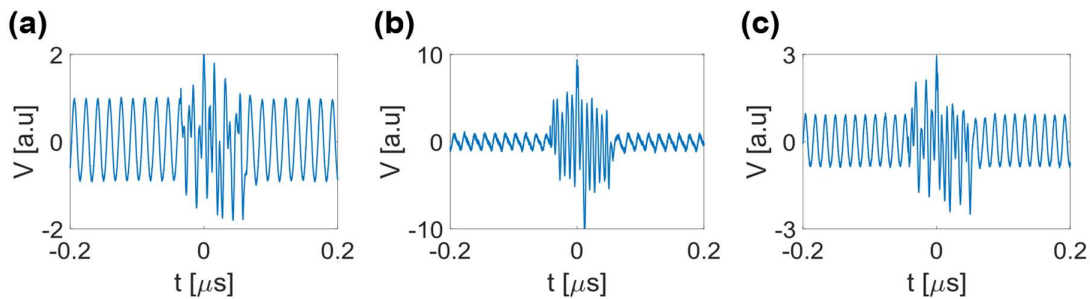

**Fig. 2. Experimental observation of time-domain BIC, with a Varactor of Q<1000.** The panels show the output signal (voltage) for a pure sinusoidal output signal at 3 different frequencies. The output signal is normalized to the base of the oscillations at $|t| \to \infty$. **(a-c)** Output signal for a sinusoidal input signal at frequencies of 57, 62 and 67 MHz, respectively. The output signal at 62 MHz, the frequency that matches the expected frequency of the time-domain BIC, displays a clear localized peak that decays on either side, as expected from a time-domain BIC. The BIC "rides" on a small residual periodic structure (pedestal). For comparison, at the two detuned frequencies (57MHz and 67MHz), the input signal is reshaped by the transmission line to a structure that displays large periodic oscillations even far away from the center.

When the varactor in our circuit is replaced with a varactor of the higher-Q, SMV1430 series, the BIC shifts upward in frequency to ≈ 76 MHz, consistent with the diode's larger tuning ratio and different C–V dependence (Fig. 3). The improved quality factor enhances the contrast between bound-state and the unbound regimes. The higher quality factor reduces resistive dissipation and parasitic leakage in the time-varying capacitance, thereby facilitating the interference condition responsible for temporal confinement to be more accurately satisfied. The systematic shift of the BIC frequency demonstrates that temporal confinement is governed by the precise functional form of C(t), which is set by the varactor's response and the applied bias waveform, rather than by the fixed spatial geometry of the circuit. As can be clearly seen in Fig. 3b, the EM power carried by the BIC, which is proportional to the voltage square $W(t) \propto |V(t)|^2$, is about 400 times larger in the time-domain BIC than in the periodic pedestal, highlighting the strong temporal localization in the frequency matching the BIC.

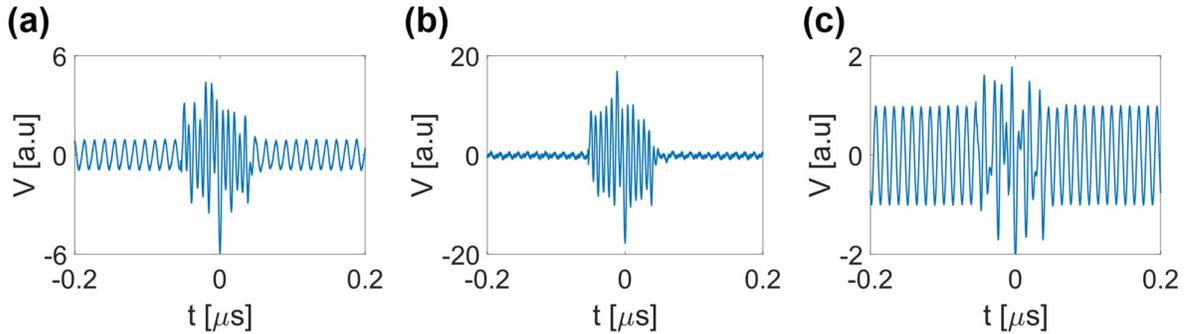

**Fig. 3. Experimental observation of time-domain BIC with a Varactor of Q ≈ 2400.** The panels show the output signal (voltage) for a pure sinusoidal output signal at 3 different frequencies. The output signal is normalized to the base of the oscillations at $|t| \to \infty$. **(a-c)** Output signal for a sinusoidal input signal at frequencies of 72, 76 and 80 MHz, respectively. The output signal at 72 MHz, the frequency that matches the expected frequency of the time-domain BIC, displays a clear localized peak that decays on either side, as expected from a time-domain BIC. With the varactor of a higher Q-factor, the localization of the time-domain BIC is much stronger, and the power at its peak proportional to the voltage squared $W(t) \propto |V(t)|^2$, is ~ 400 larger than the periodic pedestal.

The wavefunctions of the time-domain BICs, presented in Figs. 2b and 3b, highlight the importance of another unique feature: **symmetry**. Namely, the "traditional" spatial

BIC from the 1929 visionary paper(*1*) reveals that, when the underlying potential is symmetric - the spatial BIC is symmetric. On this background, the time-domain BIC presented in Figs 2 and 3 shows the opposite: the underlying modulated potential is symmetric but the wavefunction of the wave trapped by this symmetric modulation is anti-symmetric, exactly as predicted for the theoretical time-domain BIC in [52]. This feature is shown in Fig. 4, which highlights the role of symmetry: the modulation is symmetric (top panels) while both the calculated and the experimentally observed time-domain BIC (bottom panels) are anti-symmetric. Specifically in our experimental network system, while C(t) is symmetric in time, the measured voltage of the time-domain BIC state, $V_{BIC}$ (t), is anti-symmetric. This feature highlights the fundamental difference between the traditional spatial BIC and the time-domain BIC. It arises from the fact that the spatial BIC, being a solution of the Schrödinger equation, has only first-derivative in time, whereas the time-domain BIC is second order is both space and time. It is interesting to notice what happens when the input signal is not at the correct frequency, as shown in Fig. 2, panels (a), (c) and in Fig. 3, panels (a), (c). For those input signals whose frequencies do not match the BIC frequency, the output signals are neither symmetric nor ant-symmetric in time. The time-domain BIC, with its anti-symmetric wavefunction, forms only at the predesigned BIC frequency.

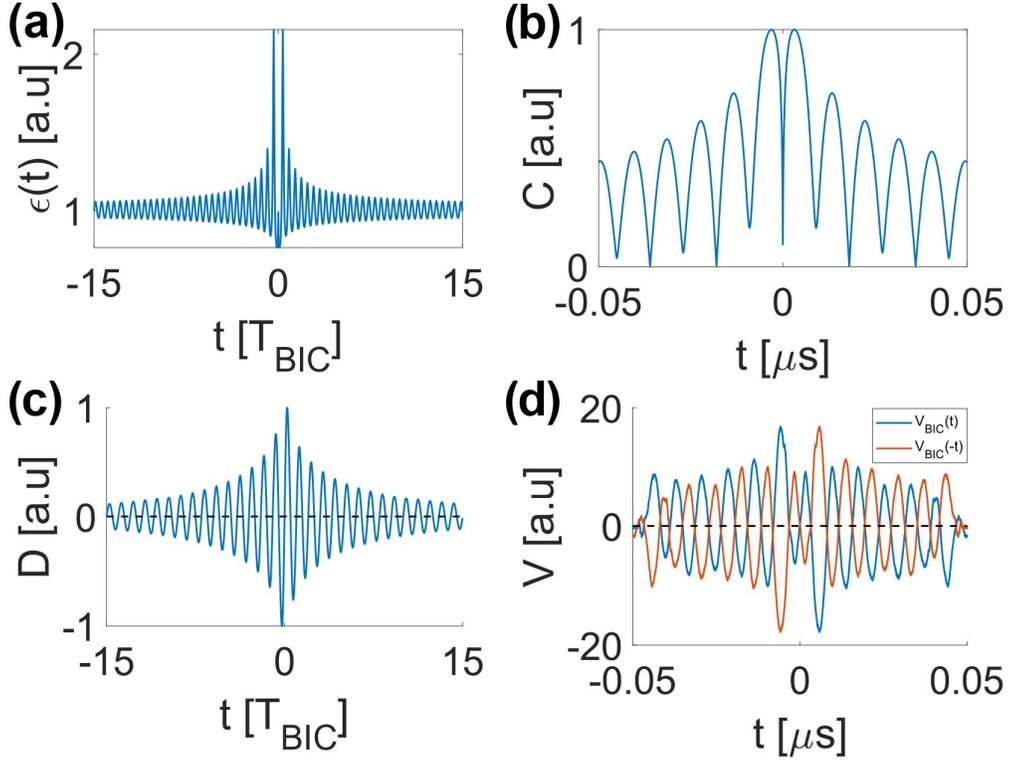

**Fig. 4. Symmetry of the calculated (left column) and experimentally realized (right column) time-domain BIC and the modulation supporting it.** (a) Calculated modulation of $\varepsilon(t)$ that produces time-domain BIC in a homogeneous dielectric medium, from (*52*). The modulation is symmetric in time. (b) Modulation applied to the varactor in the network system, showing that $C(t)$ is symmetric in time. (c) Calculated wavefunction $D(t)$ of the time-domain BIC in a homogeneous dielectric medium, from (*52*). The calculated wavefunction of the time-domain BIC is antisymmetric. (d) Measured voltage $V_{BIC}(t)$ of the time-domain BIC (blue) in our network system and its mirror image, $V_{BIC}(-t)$, (red). As shown, the measured time-domain BIC is antisymmetric while the modulation is symmetric, exactly as predicted theoretically.

In conclusion, we presented the first experimental observation of a time-domain bound state in the continuum. The time-domain BIC exhibits some similarities to its spatial counterparts, specifically in the fact that both describe wavepackets carrying finite power whose energies are embedded in the continuum of unbound states. However, there are also some fundamental differences between the spatial BIC and the time-domain BIC. Namely, the spatial BIC envisioned in 1929 is a solution of the Hermitian Schrödinger equation, whereas the time-domain BIC is created by a system that is inherently non-Hermitian, as the time-varying potential breaks time-translation symmetry and energy conservation. The differences can also be seen by examining the symmetries of the two BIC states: the time-domain BIC is anti-symmetric while the modulation is symmetric, in contrast to the original Wigner-von Neumann solution, where both the potential and the BIC have the same symmetry. This observation emphasizes the importance of symmetry in the time-domain BIC mode. The experiments presented here establish the time-domain BIC as a fundamental physical element of wave propagation. We anticipate that this work will stimulate experimental observations of time-domain BICs across diverse wave systems and motivate further research on new and exciting topics. Examples range from two-dimensional BICs in space-time systems(*53*), relating time-domain BICs to topology and especially to topological photonic time-crystals and topological space-time crystal(*49, 54*), the possibility of finding vectorial high-dimensional BIC solutions of Maxwell's equations – possibly in the form of skymionic(*55*) BICs, and time-domain BICs in the quantum regime associated with the spontaneous generation of entangled photon pairs arising from the modulation, similar to their creation in exponentially large number in photonic time-crystals(*50*).

## *Acknowledgments:*

*Funding:* Both the Purdue authors and the Technion authors gratefully acknowledge the support of the US Air Force Office of Scientific Research (AFOSR). The Technion authors acknowledge the support of the Breakthrough Program (Mapats) of the Israel Science Foundation.

# References:

1. J. von Neuman, E. Wigner, Uber merkwürdige diskrete Eigenwerte. Uber das Verhalten von Eigenwerten bei adiabatischen Prozessen. *Phys. Z.* **30**, 467–470 (1929).

2. D. C. Marinica, A. G. Borisov, S. V. Shabanov, Bound States in the Continuum in Photonics. *Phys. Rev. Lett.* **100**, 183902 (2008).

3. N. Moiseyev, Suppression of Feshbach Resonance Widths in Two-Dimensional Waveguides and Quantum Dots: A Lower Bound for the Number of Bound States in the Continuum. *Phys. Rev. Lett.* **102**, 167404 (2009).

4. X. Gao, B. Zhen, M. Soljačić, H. Chen, C. W. Hsu, Bound States in the Continuum in Fiber Bragg Gratings. *ACS Photonics* **6**, 2996–3002 (2019).

5. F. Monticone, A. Alù, Embedded Photonic Eigenvalues in 3D Nanostructures. *Phys. Rev. Lett.* **112**, 213903 (2014).

6. M. V. Rybin, K. L. Koshelev, Z. F. Sadrieva, K. B. Samusev, A. A. Bogdanov, M. F. Limonov, Y. S. Kivshar, High-$Q$ Supercavity Modes in Subwavelength Dielectric Resonators. *Phys. Rev. Lett.* **119**, 243901 (2017).

7. C. M. Linton, P. McIver, Embedded trapped modes in water waves and acoustics. *Wave Motion* **45**, 16–29 (2007).

8. A. A. Lyapina, D. N. Maksimov, A. S. Pilipchuk, A. F. Sadreev, Bound states in the continuum in open acoustic resonators. *J. Fluid Mech.* **780**, 370–387 (2015).

9. F. H. Stillinger, D. R. Herrick, Bound states in the continuum. *Phys. Rev. A* **11**, 446–454 (1975).

10. M. Robnik, A simple separable Hamiltonian having bound states in the continuum. *J. Phys. Math. Gen.* **19**, 3845 (1986).

11. E. N. Bulgakov, K. N. Pichugin, A. F. Sadreev, I. Rotter, Bound states in the continuum in open Aharonov-Bohm rings. *JETP Lett.* **84**, 430–435 (2006).

12. A. F. Sadreev, E. N. Bulgakov, I. Rotter, Bound states in the continuum in open quantum billiards with a variable shape. *Phys. Rev. B* **73**, 235342 (2006).

13. V. A. Sablikov, A. A. Sukhanov, Helical bound states in the continuum of the edge states in two dimensional topological insulators. *Phys. Lett. A* **379**, 1775–1779 (2015).

14. G. Gamow, Zur Quantentheorie des Atomkernes. *Z. Für Phys.* **51**, 204–212 (1928).

15. L. Fonda, R. G. Newton, Theory of resonance reactions. *Ann. Phys.* **10**, 490–515 (1960).

16. N. Moiseyev, Quantum theory of resonances: calculating energies, widths and cross-sections by complex scaling. *Phys. Rep.* **302**, 212–293 (1998).

17. S. Fan, W. Suh, J. D. Joannopoulos, Temporal coupled-mode theory for the Fano resonance in optical resonators. *JOSA A* **20**, 569–572 (2003).

18. N. Moiseyev, *Non-Hermitian Quantum Mechanics* (Cambridge University Press, 2011).

19. M.-S. Hwang, H.-C. Lee, K.-H. Kim, K.-Y. Jeong, S.-H. Kwon, K. Koshelev, Y. Kivshar, H.-G. Park, Ultralow-threshold laser using super-bound states in the continuum. *Nat. Commun.* **12**, 4135 (2021).

20. F. Capasso, C. Sirtori, J. Faist, D. L. Sivco, S.-N. G. Chu, A. Y. Cho, Observation of an electronic bound state above a potential well. *Nature* **358**, 565–567 (1992).

21. Y. Plotnik, O. Peleg, F. Dreisow, M. Heinrich, S. Nolte, A. Szameit, M. Segev, Experimental Observation of Optical Bound States in the Continuum. *Phys. Rev. Lett.* **107**, 183901 (2011).

22. C. W. Hsu, B. Zhen, A. D. Stone, J. D. Joannopoulos, M. Soljačić, Bound states in the continuum. *Nat. Rev. Mater.* **1**, 16048 (2016).

23. C. W. Hsu, B. Zhen, J. Lee, S.-L. Chua, S. G. Johnson, J. D. Joannopoulos, M. Soljačić, Observation of trapped light within the radiation continuum. *Nature* **499**, 188–191 (2013).

24. S. Weimann, Y. Xu, R. Keil, A. E. Miroshnichenko, A. Tünnermann, S. Nolte, A. A. Sukhorukov, A. Szameit, Y. S. Kivshar, Compact Surface Fano States Embedded in the Continuum of Waveguide Arrays. *Phys. Rev. Lett.* **111**, 240403 (2013).

25. A. Cerjan, C. W. Hsu, M. C. Rechtsman, Bound States in the Continuum through Environmental Design. *Phys. Rev. Lett.* **123**, 023902 (2019).

26. A. Cerjan, M. Jürgensen, W. A. Benalcazar, S. Mukherjee, M. C. Rechtsman, Observation of a Higher-Order Topological Bound State in the Continuum. *Phys. Rev. Lett.* **125**, 213901 (2020).

27. P. S. Pankin, B.-R. Wu, J.-H. Yang, K.-P. Chen, I. V. Timofeev, A. F. Sadreev, One-dimensional photonic bound states in the continuum. *Commun. Phys.* **3**, 91 (2020).

28. L. Huang, Y. K. Chiang, S. Huang, C. Shen, F. Deng, Y. Cheng, B. Jia, Y. Li, D. A. Powell, A. E. Miroshnichenko, Sound trapping in an open resonator. *Nat. Commun.* **12**, 4819 (2021).

29. Z. F. Sadrieva, M. A. Belyakov, M. A. Balezin, P. V. Kapitanova, E. A. Nenasheva, A. F. Sadreev, A. A. Bogdanov, Experimental observation of a symmetry-protected bound state in the continuum in a chain of dielectric disks. *Phys. Rev. A* **99**, 053804 (2019).

30. B. Zhen, C. W. Hsu, L. Lu, A. D. Stone, M. Soljačić, Topological Nature of Optical Bound States in the Continuum. *Phys. Rev. Lett.* **113**, 257401 (2014).

31. W. A. Benalcazar, A. Cerjan, Bound states in the continuum of higher-order topological insulators. *Phys. Rev. B* **101**, 161116 (2020).

32. E. N. Bulgakov, D. N. Maksimov, Topological Bound States in the Continuum in Arrays of Dielectric Spheres. *Phys. Rev. Lett.* **118**, 267401 (2017).

33. B.-J. Yang, M. Saeed Bahramy, N. Nagaosa, Topological protection of bound states against the hybridization. *Nat. Commun.* **4**, 1524 (2013).

34. F. R. Morgenthaler, Velocity Modulation of Electromagnetic Waves. *IRE Trans. Microw. Theory Tech.* **6**, 167–172 (1958).

35. J. T. Mendonça, P. K. Shukla, Time Refraction and Time Reflection: Two Basic Concepts. *Phys. Scr.* **65**, 160 (2002).

36. A. M. Shaltout, M. Clerici, N. Kinsey, R. Kaipurath, J. Kim, E. G. Carnemolla, D. Faccio, A. Boltasseva, V. M. Shalaev, M. Ferrera, "Doppler-Shift Emulation Using Highly Time-Refracting TCO Layer" in *Conference on Lasers and Electro-Optics (2016), Paper FF2D.6* (Optica Publishing Group, 2016; https://opg.optica.org/abstract.cfm?uri=CLEO_QELS-2016-FF2D.6), p. FF2D.6.

37. Y. Zhou, M. Z. Alam, M. Karimi, J. Upham, O. Reshef, C. Liu, A. E. Willner, R. W. Boyd, Broadband frequency translation through time refraction in an epsilon-near-zero material. *Nat. Commun.* **11**, 2180 (2020).

38. H. Moussa, G. Xu, S. Yin, E. Galiffi, Y. Ra'di, A. Alù, Observation of temporal reflection and broadband frequency translation at photonic time interfaces. *Nat. Phys.* **19**, 863–868 (2023).

39. T. R. Jones, A. V. Kildishev, M. Segev, D. Peroulis, Time-reflection of microwaves by a fast optically-controlled time-boundary. *Nat. Commun.* **15**, 6786 (2024).

40. E. Lustig, O. Segal, S. Saha, E. Bordo, S. N. Chowdhury, Y. Sharabi, A. Fleischer, A. Boltasseva, O. Cohen, V. M. Shalaev, M. Segev, Time-refraction optics with single cycle modulation. *Nanophotonics* **12**, 2221–2230 (2023).

41. O. Segal, N. Konforty, O. Schiller, M. J. Tolchin, J.-P. Maria, Y. Plotnik, M. Segev, Sub-cycle time-refraction at optical frequencies. arXiv arXiv:2601.05566 [Preprint] (2026). https://doi.org/10.48550/arXiv.2601.05566.

42. A. Akbarzadeh, N. Chamanara, C. Caloz, Inverse prism based on temporal discontinuity and spatial dispersion. *Opt. Lett.* **43**, 3297–3300 (2018).

43. V. Pacheco-Peña, N. Engheta, Temporal aiming. *Light Sci. Appl.* **9**, 129 (2020).

44. H. Li, S. Yin, E. Galiffi, A. Alù, Temporal Parity-Time Symmetry for Extreme Energy Transformations. *Phys. Rev. Lett.* **127**, 153903 (2021).

45. O. Schiller, Y. Plotnik, G. Bartal, M. Segev, Negative Index Makes a Perfect Time-Domain Lens, Generating Slow Playback of Ultrafast Events. arXiv arXiv:2512.03985 [Preprint] (2025). https://doi.org/10.48550/arXiv.2512.03985.

46. J. R. Zurita-Sánchez, P. Halevi, J. C. Cervantes-González, Reflection and transmission of a wave incident on a slab with a time-periodic dielectric function $\epsilon(t)$. *Phys. Rev. A* **79**, 053821 (2009).

47. J. R. Reyes-Ayona, P. Halevi, Observation of genuine wave vector (k or β) gap in a dynamic transmission line and temporal photonic crystals. *Appl. Phys. Lett.* **107**, 074101 (2015).

48. F. Biancalana, A. Amann, A. V. Uskov, E. P. O'Reilly, Dynamics of light propagation in spatiotemporal dielectric structures. *Phys. Rev. E* **75**, 046607 (2007).

49. E. Lustig, Y. Sharabi, M. Segev, Topological aspects of photonic time crystals. *Optica* **5**, 1390–1395 (2018).

50. M. Lyubarov, Y. Lumer, A. Dikopoltsev, E. Lustig, Y. Sharabi, M. Segev, Amplified emission and lasing in photonic time crystals. *Science* **377**, 425–428 (2022).

51. A. Dikopoltsev, Y. Sharabi, M. Lyubarov, Y. Lumer, S. Tsesses, E. Lustig, I. Kaminer, M. Segev, Light emission by free electrons in photonic time-crystals. *Proc. Natl. Acad. Sci.* **119**, e2119705119 (2022).

52. O. Schiller, Y. Plotnik, O. Segal, M. Lyubarov, M. Segev, Time-Domain Bound States in the Continuum. *Phys. Rev. Lett.* **133**, 263802 (2024).

53. Y. Sharabi, A. Dikopoltsev, E. Lustig, Y. Lumer, M. Segev, Spatiotemporal photonic crystals. *Optica* **9**, 585–592 (2022).

54. O. Segal, Y. Plotnik, E. Lustig, Y. Sharabi, M.-I. Cohen, A. Dikopoltsev, M. Segev, Two-Dimensional Topological Edge States in Periodic Space-Time Interfaces. *Phys. Rev. Lett.* **135**, 163801 (2025).

55. S. Tsesses, E. Ostrovsky, K. Cohen, B. Gjonaj, N. H. Lindner, G. Bartal, Optical skyrmion lattice in evanescent electromagnetic fields. *Science* **361**, 993–996 (2018).

56. K. O'Brien, C. Macklin, I. Siddiqi, X. Zhang, Resonant Phase Matching of Josephson Junction Traveling Wave Parametric Amplifiers. *Phys. Rev. Lett.* **113**, 157001 (2014).